\documentclass[12pt]{article}

\usepackage[english]{babel}
\usepackage{amsfonts}
\usepackage{amssymb}
\usepackage{graphicx}
\newcommand{\be}{\begin{equation}}
\newcommand{\ee}{\end{equation}}
\newcommand{\bea}{\begin{eqnarray}}
\newcommand{\eea}{\end{eqnarray}}

\begin{document}

\begin{titlepage}

\begin{flushright}
{\tt
    hep-th/0512180}
 \end{flushright}

\bigskip

\begin{center}

{\Large \bf{Quantum black holes and holography }}

\bigskip
\bigskip\bigskip
 Alessandro  Fabbri\footnote{afabbri@ific.uv.es}\footnote{Talk given
 at the Conference ``Constrained dynamics and quantum gravity QG05", 
 Cala Gonone (Italy), September 2005}

\end{center}

\bigskip%

\footnotesize \noindent {\it Departamento de F\'{\i}sica
Te\'orica and
    IFIC, Centro Mixto Universidad de Valencia-CSIC.
    Facultad de F\'{\i}sica, Universidad de Valencia,
        Burjassot-46100, Valencia, Spain.}

\bigskip

\bigskip

\begin{center}
{\bf Abstract}
\end{center}
It is technically difficult (if not impossible) to write down and
solve self-consistently the semiclassical Einstein equations in
the case of evaporating black holes. These difficulties can in
principle be overcome in an apparently very different context, the
Randall-Sundrum braneworld models in Anti-de Sitter space. Use of
Maldacena's AdS/CFT correspondence led us to formulate a
holographic conjecture for black holes localised on a brane, for
which 4D quantum corrected black holes are dual to classical 5D
black holes. This duality is applied to the computation of the
correction to the newtonian potential on the brane, with new
results on the semiclassical side, and a prediction about the
existence of static large mass braneworld black holes is made.



\end{titlepage}

\newpage

\section{Introduction}

One of the most urgent problems a self-consistent quantum gravity
theory will have to solve is a proper understanding of the quantum
properties of black holes. In the standard approach one uses the
semiclassical theory, in which matter fields are quantized in a
classical curved background (see for instance \cite{libri}).  In
this context the Einstein equations are replaced by the
semiclassical ones
\begin{equation}\label{semeinst}
G_{\mu\nu}=8\pi\langle T_{\mu\nu}\rangle\ ,
\end{equation}
where the source term is provided by the expectation values of the
stress energy tensor operator for the matter fields in a suitably
defined quantum state. The problem with these equations is that
the exact expression of $\langle T_{\mu\nu}\rangle$ for an
arbitrary geometry, needed to solve self-consistently Eqs.
(\ref{semeinst}), is not known in four dimensions. What one
usually does is to evaluate $\langle T_{\mu\nu}\rangle$ in a fixed
classical background, such as for instance the Schwarzschild
spacetime (but even in this case one has to resort to
approximations), and then use it as source in (\ref{semeinst}) to
compute the $O(\hbar)$ corrections to the classical geometry. This
provides a good approximation to the exact semiclassical solution
when the quantum terms are small compared to the classical ones.
Unfortunately, in the physically most interesting situation, the
black holes evaporation process \cite{hawk}, the fixed background
approximation cannot always be trusted. In particular, it looses
validity at the late stages of the evolution, and this prevents us
to make quantitative predictions about the fate of quantum black
holes. One might object that when the black hole reaches the
Planck mass a full quantum gravitational treatment is needed to
properly address this issue. To find a way out to this problem one
usually argues that quantum gravitational effects should always be
negligible compared to those due to a large number $N$ of matter
fields.

In this (almost) hopeless situation, three and a half years ago a
new framework was proposed \cite{efktan} which allows in principle
to address this difficult issue in an apparently very different
context, namely braneworld models in Anti-de Sitter space (in
particular, the Randall-Sundrum model RS2 \cite{rs2}). The tool
used is the so called holographic interpretation in AdS
braneworlds, which results from the application of Maldacena's
AdS/CFT correspondence \cite{adscft} to RS2. This is briefly
reviewed in Section 2. Application to the search of black holes
localised on the brane (Section 3) is quite straightforward, and
leads to the exciting possibility that the quantum properties of
4D black holes can be understood by solving 5D classical Einstein
equations. An important verification of the holographic
interpretation in AdS braneworlds is the computation of the
correction to the Newtonian potential $\phi$ on the brane (Section
4), which from the AdS/CFT perspective can be calculated in two
very different, but equivalent ways, i.e. classical in 5D
\cite{gata} and quantum mechanical in 4D \cite{duli}, both in the
weak field limit. Using the semiclassical framework Eqs.
(\ref{semeinst}) (Sections 5 and 6) we shall see how, via a
numerical computation of $\langle T_{\mu\nu}\rangle$ in the
Schwarzschild spacetime for matter fields in the (zero
temperature) Boulware state, to rederive the result for $\phi$ (a
nontrivial computation indeed) from the asymptotic quantum
corrected Schwarzschild geometry. Moreover, since our approach
allows to go beyond the weak field limit we shall  make a
prediction about the existence and the spacetime structure of
large mass black holes in AdS braneworlds. Finally, in Section 7
we end by highlighting that this new framework gives us the way to
overcome the technical difficulties in the semiclassical theory to
deal with evaporating black holes.

\section{Holographic interpretation in AdS \\ braneworlds}

The AdS/CFT duality \cite{adscft} predicts a one-to-one
correspondence between a quantum gravity theory defined in Anti-de
Sitter space and a Conformal Field Theory living in its boundary
at infinity. The relevant duality we shall be interested in is
between Type IIB String Theory in $AdS_5 \times S^5$ (which arises as
the near-horizon limit of $N$ coincident $D3$-branes) and
${\cal{N}=} 4$ $SU(N)$ super Yang-Mills theory. At leading order
(large $N$ or, equivalently, large $AdS$ length $L$) the
correspondence is between a classical theory in the bulk and a
strongly coupled CFT in the 't Hooft planar limit $\lambda\equiv
g_{YM}^2 N \gg 1$. To apply this formalism to AdS braneworlds we
shall need a further ingredient, which is provided by the so
called UV-IR connection \cite{suwi}. By expressing AdS metric in
Poincar\'e coordinates
\begin{equation}\label{rsvacuum}
ds^2=\frac{L^2}{z^2}\left[ \eta_{\mu\nu}dx^\mu dx^\nu +dz^2
\right]\ ,
\end{equation}
where (timelike) infinity corresponds to $z=0$, an infrared cutoff
in (\ref{rsvacuum}) at $z=\delta$ corresponds to a UV cutoff in
the dual CFT at distances $\delta$. \\ On the other hand, in the
Randall-Sundrum model RS2 \cite{rs2} our universe (the brane) is
identified with a hypersurface in five dimensional AdS space (the
bulk). In the language of modified AdS/CFT (called cutoff AdS/CFT
\cite{cutoffadscft}) the brane becomes the boundary of AdS space
and its position fixes the cutoff scale $\delta=L$. In addition,
by studying the linear gravitational perturbations $h_{\mu\nu}$
around the Randall-Sundrum vacuum (\ref{rsvacuum}) one finds that
the dependence on the extra-dimensional coordinate $z$ is
expressed via the following Schr\"odinger type equation (the brane
is now placed at $z=0$)
\begin{equation}\label{pertgrav}
\left[-\partial_z^2 + V(z) \right]h_{\mu\nu} = m^2 h_{\mu\nu} \ ,
\end{equation}
where $V(z)=\frac{15}{8L^2(|z|/L+1)^2}-\frac{3}{2L}\delta(z)$ is
called the Minkowski volcano potential. The zero mode $m=0$ gets
trapped on the brane, reproducing 4D gravity there, and moreover
there is a continuum of Kaluza-Klein modes with no gap. \\ To sum
up the results presented in this section, we have seen that the
RS2 model provides a realization of the cutoff AdS/CFT
correspondence with the brane, our universe, playing the role of
the boundary of AdS space. In the brane we have both the dual
cutoff CFT and Einstein gravity. In the large $N$ (or $L$) limit
this leads to the formulation of the holographic interpretation in
RS braneworlds, which states that {\it the dual of the classical
bulk theory is a (strongly coupled) quantum CFT in the planar
limit coupled to classical gravity}. In this view the bulk KK
modes are dual to the brane CFT modes. We will see a concrete
realization of this last duality in the computation of the
correction to the Newtonian potential in Section 4.

\section{The holographic conjecture for black holes on the brane}

It is now quite straightforward to apply the formalism just
presented to the search of black hole solutions in AdS braneworlds
\cite{efktan}: {\it 4D black holes (with large mass) localised on
the brane found by solving the classical bulk equations in $AdS_5$
are quantum corrected black holes and not classical ones}. By
quantum corrected black holes we mean, of course, solutions to the
semiclassical Einstein equations (\ref{semeinst}). This conjecture
has great appeal, as it allows to use known semiclassical results
to predict the behavior of classical 5D solutions (see Section 6)
and the other way around, i.e. use of the bulk classical Einstein
equations to overcome the technical difficulties due to the
incomplete knowledge of Eqs. (\ref{semeinst}). \\ We must warn,
however, that the semiclassical results are mostly derived from
the assumption that the matter fields are weakly coupled (free),
whereas in this context the CFT is strongly coupled.  Despite this
fact, there is an excellent agreement between the exact solutions
for black holes localised on a 2 brane \cite{ehm} and the 2+1
semiclassical solutions describing quantum corrected conical
singularities and quantum corrected BTZ black holes (see for
instance \cite{twoplusone}). \\ In the most interesting case of
$AdS_5$ with a 3 brane, in \cite{bgm} the Oppenheimer-Snyder model
(collapsing dust) on the brane was considered. Unlike in standard
4D gravity, where due to Birkhoff's theorem the interior solution
for the star can be smoothly matched to a unique exterior
represented by Schwarzschild, the effective equations on the brane
do not allow the matching with a static exterior. Morever, despite
many efforts \cite{molti, kutana} it hasn't been possible to
find static black hole solutions to the bulk equations. One
possible explanation is of course that the problem is technically
difficult. However, these results can be naturally explained in
terms of the holographic conjecture for black holes on the brane:
due to the inclusion of quantum effects black holes cannot be
static, as they evaporate via the {\it Hawking effect}. However,
it is important to stress that no actual proof of it has been
provided so far in $AdS_5$.

\section{Correction to the Newtonian potential $\phi$: 5D classical vs.
4D quantum computation}

An important verification of the holographic interpretation in AdS
braneworlds concerns the computation of the correction to the
Newtonian potential on the brane. This was computed classically in
$AdS_5$ in \cite{gata} with the result
\begin{equation}\label{corrnewbulk}
\phi = \frac{M}{r}(1+\frac{2}{3}\frac{L^2}{r^2})\ ,
\end{equation} where the Newtonian term comes from the zero mode
in the decomposition (\ref{pertgrav}) and the additional
contribution is derived by integrating over all the continuum KK
modes. Since these latter are supposed to correspond to the modes
of the dual CFT, this same result must be derivable in a 4D
quantum context by computing the one graviton exchange diagram
with the insertion of matter loops \cite{donoghue}. In the case of
conformal fields the calculation was performed long ago by Duff
\cite{duff}
\begin{equation}\label{quantcorrphi}
\phi= \frac{M}{r}(1+ \frac{\alpha}{r^2})\ , \end{equation} where
the coefficient $\alpha$ depends on the spin of the field
\begin{equation}
\alpha=\frac{1}{45\pi}(12 N_1 + +3 N_{1/2} + N_0)\ .
\end{equation}
We observe that the correction term ($1/r^3$) decays in the same
way as in (\ref{corrnewbulk}), but to show complete equivalence
between the two results one needs to check the numerical
coefficients. This can be done by specifying the matter content of
the CFT, i.e. $N_1=N^2$, $N_{1/2}=4N^2$ and $N_0=6N^2$ for
${\cal{N}}=4$ SU(N) SYM, plus the relation $N^2=\pi L^2$ (derived
from AdS/CFT and the RS relation between 4D and 5D Newton's
constants). Summing the contribution of all the fields involved
(note that this is a weak coupling calculation) the result
(\ref{corrnewbulk}) is exactly reproduced \cite{duli}.

\section{Computation of $\phi$ via the 4D backreaction equations}

We shall now see how nontrivial is to reproduce the result
(\ref{quantcorrphi}) in the semiclassical context. The idea is to
read off $\phi$ directly from the quantum corrected Schwarzschild
solution. To solve Eqs. (\ref{semeinst}) we need an expression for
$\langle T_{\mu\nu}\rangle $ in the Schwarzschild spacetime for,
say, a conformal scalar field (only the leading order term is
important). \\ The first problem we face is: what is the correct
choice for the quantum state of the matter field? We know we have
three physically distinct possibilities. The first is the Boulware
state \cite{boul}, constructed by requiring that at infinity it
reduces to Minkowski ground state. In this state $\langle
T_{\mu\nu}\rangle$ vanishes at large $r$, but is strongly
divergent at the horizon. It is possible to construct a state, the
Hartle-Hawking state \cite{haha}, regular at the horizon, but
again we pay a price: $\langle T_{\mu\nu}\rangle$ is nonvanishing
asymptotically, where it describes thermal radiation at the
Hawking temperature $T_H=1/8\pi M$. Physically, this is a black
hole in thermal equilibrium with its own radiation. Finally, the
most interesting case is the Unruh state \cite{un}, which
reproduces the late-time properties of the radiation emitted in
the process of black hole formation (Hawking radiation). Clearly,
only the Boulware vacuum gives a semiclassical configuration which
is static and asymptotically flat (Hartle-Hawking gives a static
non asymptotically flat solution and Unruh an asymptotically flat
time dependent one). This is indeed what we need for our
purposes. \\
Let us now look at the analytic approximations proposed in the
literature for $\langle T_{\mu\nu}\rangle$ in Boulware state to
check the leading order behavior. Brown and Ottewill \cite{brot}
and Frolov and Zelnikov \cite{frozel} derived two distinct
expressions valid only for the conformal case, the first using the
transformation properties of the stress tensor in two conformally
related spacetimes and the second, the Killing
approximation, writing down the most general tensor with the
correct trace anomaly via the Killing vector, the
curvature and their derivatives up to a certain order.
Anderson, Hiscock and Samuel \cite{anhisa} developed an
approximation valid for all couplings $\xi$ with the scalar
curvature based on the WKB approximation.
They all agree that the leading term  is
\begin{equation}\label{leadanapp}
\langle T_{\mu\nu}\rangle \sim O(M^2/r^6)\ ,
\end{equation}
which was considered to be a reasonable result since it is of the
same order as the trace anomaly. However, it is not difficult to
show that this would give a quantum correction to $\phi$ of the
order $M^2/r^4$ and not $M/r^3$ as in (\ref{quantcorrphi})! \\
A numerical computation was then performed in \cite{abf} till
large values of $r$ based on the method developed in
\cite{anhisa}. The results of this investigation indicate that the
correct leading order term is not (\ref{leadanapp}) but rather
$M/r^5$.
Moreover, fitting the numerical results up to two significant
digits we find that
\begin{eqnarray} \langle T^t_{\ t} \rangle &=& \frac{M}{60\pi^2
r^5} + (\xi -\frac{1}{6})\frac{M}{4\pi^2 r^5}\ , \nonumber \\
\langle T^r_{\ r} \rangle &=& \frac{M}{120\pi^2 r^5} -
 (\xi -\frac{1}{6})\frac{M}{4\pi^2 r^5}\ , \nonumber \\
\langle T^\theta_{\ \theta} \rangle &=& -\frac{M}{80\pi^2 r^5} +
(\xi -\frac{1}{6})\frac{3M}{8\pi^2 r^5}\ . \end{eqnarray} It is
now straighforward to insert the above expressions into Eqs.
(\ref{semeinst}) to find the asymptotic quantum corrected
Schwarzschild metric
\begin{equation}
ds^2\simeq -\left[1-\frac{2M}{r}(1+\frac{\alpha}{r^2})\right]dt^2
+
\left[1-\frac{2M}{r}(1+\frac{\beta}{r^2})\right]^{-1}dr^2+r^2d\Omega^2
\ ,
\end{equation}
where
\begin{equation}
\alpha=\frac{1}{45\pi}-(\xi-\frac{1}{6})\frac{1}{6\pi}\ ,\ \ \
\beta=\frac{1}{30\pi}+(\xi-\frac{1}{6})\frac{1}{2\pi}\
,\end{equation} and read from the $g_{tt}$ component the result
for $\phi$
\begin{equation}\label{quancorrphisem}\phi= \frac{M}{r}(1+\frac{\alpha}{r^2})\ .
\end{equation}
Remarkably, our result (\ref{quancorrphisem}) agrees with
(\ref{quantcorrphi}) in the conformal case $\xi=1/6$ and also
matches the known result \cite{hali} in the minimally coupled case
$\xi=0$. Of course, to completely reproduce the Randall Sundrum
result (\ref{corrnewbulk}) one would need to perform the numerical
analysis for spin 1/2 and spin 1 fields as well.

\section{Prediction about the existence of large mass static braneworld black holes}

The successful check discussed in the previous section gives us
the confidence to try to go beyond the weak field limit, which is
the major limitation of the previous methods encountered, and make
a prediction about the existence of static braneworld black holes
with large mass. It is well known that at the horizon $r_H=2M$ the
Boulware state stress tensor diverges as \cite{chrisfull}
\begin{equation} \label{divboul}
\langle T^\mu_{\ \nu}\rangle \sim
\frac{(2N_1+\frac{7}{4}N_{1/2}+N_0)}{30\ 2^{12}\pi^2 M^4f^2}(1,
-1/3,-1/3,-1/3)\ ,\ \ \ f=(1-2M/r) \ .
\end{equation}
It would be interesting to understand whether and how strong
coupling effects modify this result. \\ Naive insertion of
(\ref{divboul}) in Eqs. (\ref{semeinst}) indicates that the
quantum terms destroy the horizon and replace it with a curvature
singularity. This is confirmed by a careful numerical analysis of
the backreaction equations in the s-wave approximation
\cite{ffnos}. The quantum corrected solution is very similar to
Schwarzschild till very close to $r=2M$, but then big differences
emerge: in particular there appears a bouncing surface for the
radial function $r$ (the radius of the two-spheres), which
prevents the formation of an event horizon, and, beyond it, a
curvature singularity.  \\ But do we expect, on physical grounds,
quantum effects to be large at the horizon for a large mass black
hole? Since all curvature invariants are small the natural answer
is no, at least in the semiclassical approximation. What the
peculiar properties of the static solution just discussed indicate
is that the natural thing for a black hole is to be time dependent
and evaporate. Indeed, the usual physical interpretation of the
Boulware state is that it describes vacuum polarization due to
matter fields around a static star (whose radius is bigger  than
$2M$) and not a black hole. \\ One possible exception to this
conclusion could be if the system has charges (say, gauge charges)
allowing the presence of zero-temperature configurations. Indeed,
it has been shown in \cite{anhilo} that the stress energy tensor
is regular at the horizon of an extremal Reissner-Nordstr\"om
black hole. \\ Coming back, finally, to the dual problem in 5D we
are thus led to the following prediction: {\it either macroscopic
static braneworld black holes do not exist or, if they do, they
have zero temperature}.

\section{Open questions}

The main qualitative improvement holography is giving us in our
effort to understand the quantum properties of black holes is to
provide a well defined set of equations (five dimensional Einstein
equations with a negative cosmological constant), as opposed to
the incomplete knowledge of the exact form of Eqs.
(\ref{semeinst}). Of course its resolution in the physically
interesting cases is not an easy task, and the main results will
probably come via a numerical analysis. \\ Kudoh, Tanaka and
Nakamura \cite{kutana} gave numerical evidence for the existence
of small static 5D black holes ($r_H\stackrel{<}{\sim}L$), but not
of large ones. According to the holographic conjecture the
deviations from staticity should be explained in terms of Hawking
radiation in the dual theory. A check of this claim will give
strong quantitative support to the conjecture. \\
The most important application is to solve the full 5D equations
for the time dependent problem corresponding to gravitational
collapse on the brane. This is a hard problem, but, potentially, with
a high reward: to understand the details of black holes
evaporation and, possibly, to have some clues concerning the
information loss paradox. 

\bigskip \noindent
{\bf Acknowledgements:}
\noindent
It is a pleasure to thank my collaborators Paul Anderson, Roberto
Balbinot, Roberto Emparan, Nemanja Kaloper and Jos\'e
Navarro-Salas for many enlightening and stimulating
discussions.

\newpage

\smallskip

\end{document}